\documentclass[twocolumn,showpacs,preprintnumbers,amsmath,amssymb, prl]{revtex4}

\usepackage{amsmath}
\usepackage{amsfonts}
\usepackage{graphics}
\usepackage{graphicx}
\usepackage{epsfig}
\usepackage[latin1]{inputenc}

\begin{document}    
  
\preprint{LMU-ASC 31/06}

\title{Exclusion Processes with Internal States}

\author{Tobias Reichenbach, Thomas Franosch, and Erwin Frey}

\affiliation{Arnold Sommerfeld Center for Theoretical Physics (ASC) and
  Center for NanoScience (CeNS), Department of Physics,
  Ludwig-Maximilians-Universit\"at M\"unchen, Theresienstrasse 37,
  D-80333 M\"unchen, Germany}

\date{August 8, 2006}
 
\begin{abstract}
  We introduce driven exclusion processes with internal states that
  serve as generic transport models in various contexts, ranging from
  molecular or vehicular traffic on parallel lanes to spintronics.
  The ensuing non-equilibrium steady states are
  controllable by boundary as well as bulk rates. A striking
  polarization phenomenon accompanied by domain wall motion and  delocalization
  is discovered within a mesoscopic scaling. We quantify
  this observation within an analytic description providing exact phase diagrams.
  Our results are confirmed by stochastic simulations.
\end{abstract}
 

\maketitle 
   
Understanding the physical principles governing non-equilibrium
transport in one-dimensional (1D) systems has been the subject of
recent interest in both biological physics
\cite{hirokawa-1998-279, rustom-2004-303} and mesoscopic
quantum systems \cite{zutic-2004-76}. Though there are fundamental
differences due to quantum coherence effects, there is a variety of
common themes. One of them is the control of a non-equilibrium
steady state through particle injection and extraction at the
boundaries and coupling to some external field in bulk. For
example, a generic spintronic scheme \cite{zutic-2004-76} consists
of sources and drains for spin injection and extraction where the
spin orientation is controlled by a tunable effective magnetic
field. The analog in non-equilibrium statistical mechanics is the
asymmetric simple exclusion process with open
boundaries \cite{derrida-1998-301}.  In its simplest version,
particles interacting only with hard-core repulsion move
uni-directionally from the left to the right boundary, which are
acting as sources and drains, respectively. These particles may
either correspond to molecular engines like mRNA or kinesin moving
actively along molecular tracks~\cite{Howard} or to macromolecules
driven through nanoscale pores or channels~\cite{chou-1999-82}
by some external field. Already this simplest
conceivable model for collective transport exhibits phase
transitions between different non-equilibrium steady states
controlled by the entrance and exit rates at the boundaries
\cite{krug-1991-76}. It has recently been noted that a minimal
model for intracellular transport has to account for the fact that
molecular motors may enter or leave the track not only at the
boundaries but also in    
bulk~\cite{klumpp-2001-87,parmeggiani-2003-90}.  Then, the bulk
reservoir of particles acts as a gate, which can induce phase
separation \cite{parmeggiani-2003-90} in a mesoscopic limit where
the residence time of the particles is of the same order as the
transit time.   
     
\begin{figure}[htbp]
\begin{center}
\includegraphics[scale=1]{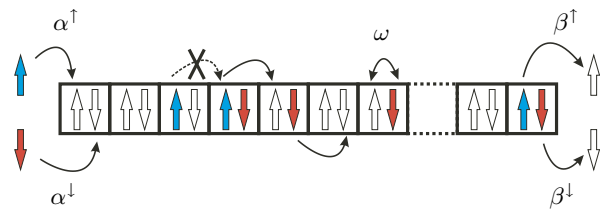}
\caption{(color online) Illustration of an exclusion model with two internal
states. Particles in states $\uparrow$ ($\downarrow$) enter with
rates $\alpha^{\uparrow}$ ($\alpha^\downarrow$), move
unidirectionally to the right within the lattice, may flip at rate
$\omega$, and leave the system at rates $\beta^{\uparrow}$
($\beta^\downarrow$), always respecting Pauli's exclusion principle.
\label{cartoon}}
\end{center}
\end{figure}

In this Letter, we introduce a generalization of the (totally)
asymmetric simple exclusion process (TASEP) in which particles
possess some discrete internal states.  For illustration, we
restrict the discussion to two states which are referred to as
spin-up ($\uparrow$) and spin-down ($\downarrow$); see
Fig.~\ref{cartoon}. Inspired by Pauli's exclusion principle, we
allow multiple occupancy of sites only if particles are in
different internal states.  The resulting dynamics underlies
surprisingly diverse situations.  The internal states may
correspond to distinct states of a molecular engine which are
allowed to simultaneously occupy the same site of a molecular
track and mutually inhibit each others motion. In the context of
molecular~\cite{hinsch-2006-}  or vehicular~\cite{helbing-2001-73}
traffic, transport on several parallel lanes may be described by
attributing internal states to vehicles (molecular motors) moving
on a single lane. Even further, considering hopping transport in
chains of quantum dots \cite{hahn-1998-73} with applied voltage, the states
are directly identified
with the spin of the electron. In this situation, the model
specified in more detail below maps to a quasi-classical version
of a non-equilibrium Ising spin chain with nearest neighbor
hopping and spin flips, where particles still respect Pauli's
exclusion principle but phase coherence is lost.

Motivated by this broad range of possible applications, we consider the
following 1d lattice model with $L$ sites and open boundaries
illustrated in Fig.~\ref{cartoon}.
At the left, particles with spin-up (-down) are injected at rate
$\alpha^\uparrow$ ($\alpha^\downarrow$), respecting Pauli's
exclusion principle, i.e. each site might at most be occupied by
one spin-up and one spin-down state. 
Within  bulk, particles hop
unidirectionally to the right at rate $r$ and may flip their spin
orientation with rate $\omega$, again under the constraint of
Pauli's exclusion principle.
Finally, the rates $\beta^{\uparrow,
\downarrow}$ control spin extraction at the right boundary. When
interpreting our model as two-lane traffic~\cite{helbing-2001-73},
the states correspond to lanes such that each site might at most
be occupied once, and spin flip translates to switching between
lanes.

The quantities of interest are the state-resolved density
$\rho^{\uparrow,\downarrow}(x)$ and current profiles
$j^{\uparrow,\downarrow}(x)$ in the non-equilibrium steady state as 
functions of the spatial
position $x$, emerging from the interplay between external
driving, coupling between the internal states (spin flips), and the
exclusion principle. We identify a cooperative transition between
a homogeneous state, where the density profiles of both internal
states exhibit no significant spatial dependence, to a
``polarized'' state, where the density profile of one state
changes abruptly along the track. The latter implies that the
polarization $\rho^\uparrow(x) - \rho^\downarrow(x)$ switches from
a ``low'' to a ``high'' value at a well-defined position
$x_\text{w}$; i.e.,  a domain wall forms. We find that this
position can be tuned by changing the rates of spin injection
(source) and extraction (drain) as well as the spin flip rate
(gate). For the transition from a polarized to
an unpolarized state  two scenarios emerge. Either the domain wall
leaves the system continuously through the left or right boundary
or it exhibits a delocalization transition. There are thus two
genuinely distinct modes of switching the polarized state of the
system on and off. 

The system's dynamics is governed by boundary ($\alpha, \beta$)
and bulk ($r, \omega$) rates, where from now on we fix the time
scale by setting the hopping rate to unity, $r=1$.  Obviously, if
the spin flip rate $\omega$ is fast, i.e., comparable with the
hopping rate, the spin degrees of freedom can be considered as
relaxed such that the system's behavior is qualitatively the same
as for  TASEP without internal states \footnote{The physics in
this limit has previously been
  discussed in terms of a two-lane model \cite{pronina-2004-37}.}.
However, if the typical number of spin flips that a particle
performs while traversing  the system becomes comparable to the
entrance and exit rates,
 one expects the
interplay between these processes to yield interesting collective
effects. In order to highlight this dynamic regime, we introduce
the gross spin flip rate $\Omega = L \omega$, which we choose to be
of the same order as the boundary rates. A proper mesoscopic
limit \cite{parmeggiani-2003-90} is defined by keeping $\Omega$
fixed as the number of lattice sites $L$ tends to infinity.

Much of the system's behavior can already be inferred   on the
basis of symmetries and current conservation. Consider the motion
of holes with a given internal state, i.e., the absence of a
particle with opposite spin. These holes move from right to left
and  flip their spin state at the same rate $\omega$. The dynamics
exhibits a particle-hole symmetry: changing the notion of
particles and holes with simultaneous interchange of ``left'' and
``right'' as well as $\alpha^{\uparrow,\downarrow} \leftrightarrow
\beta^{\downarrow,\uparrow}$ leaves the system's dynamics
invariant. In addition, there is a spin symmetry as seen by
interchanging the spin states and the injection and extraction
rates, $\alpha^\uparrow \leftrightarrow \alpha^\downarrow$ and
$\beta^\uparrow \leftrightarrow \beta^\downarrow$. Let us now
study the spin and particle currents passing through site $i$,
denoted by $j^{\uparrow, \downarrow}_i$ and $J_i=j^\uparrow_i +
j^\downarrow_i$, respectively. Since particles are not allowed to
leave or enter the system, except for the boundaries, the particle
current $J_i$ is strictly conserved and thus spatially constant,
$J=J_i$. Unlike the particle current, the spin currents are not
conserved individually. Because of spin flip processes, there is a
leakage current from one spin state to the other. Since spin flips
typically occur on time scales comparable to the time a particle
needs to traverse the system (mesoscopic limit) this 
leakage current is only weak, and consequently both spin currents
exhibit a slowly varying spatial dependence.

Similar to  TASEP, the particle current is limited either by the
left or right boundary or the capacity of the bulk. The latter
restricts the current to values below a maximal value of $1/2$.
For the left boundary, the current can not exceed $J_{\text{IN}} =
\alpha^\uparrow (1-\alpha^\uparrow) + \alpha^\downarrow
(1-\alpha^\downarrow)$, while the right boundary constrains it to
a value not larger than $J_{\text{EX}} = \beta^\uparrow
(1-\beta^\uparrow) + \beta^\downarrow (1-\beta^\downarrow)$. If
the current is below the maximal current, it is determined by the
smaller of the boundary currents: $J= \text{min} (J_{\text{IN}}, J_{\text{EX}})$
\footnote{Strictly speaking, this holds only for rates smaller than
$1/2$. Larger rates effectively act as $1/2$ \cite{reichenbach-long}.}. Depending on which of both cases
applies, we discern two complementary regions in the 
five-dimensional parameter space spanned by
$\alpha^{\uparrow,\downarrow}, \beta^{\uparrow,\downarrow}$ and
$\Omega$. We refer to the region where the total current is given
by $J_{\text{IN}}$ as the injection
  dominated region (IN), while the case $J=J_{\text{EX}}$ corresponds to the
extraction dominated region (EX). Note that both regions
are connected by particle-hole symmetry. Since this symmetry is
discrete, we expect discontinuous phase transitions upon
crossing the border from the injection to the extraction dominated
region, i.e. the IN-EX boundary.  At this boundary the
system exhibits phase coexistence, which similar to TASEP
\cite{macdonald-1968-6} manifests itself in a delocalized domain
wall between a low (LD) and high density (HD) state of both
spin states. Thus, based on mere symmetry arguments, we conclude
that across the IN-EX boundary a delocalization
transitions appears.  The simultaneous formation of a domain wall
in the density profiles of both spin states exclusively
occurs at the IN-EX boundary. This restriction also
originates in the conserved particle current, as can be seen by
the following argument: For the presence of a domain wall in the
spin-up density profile, the spin-up current in the vicinity of
the left boundary
is determined by the entrance rate,
$j^\uparrow_{i=1}=\alpha^\uparrow(1-\alpha^\uparrow)$, while at
the right boundary the exit rate specifies its value to
$j^\uparrow_{i=L}=\beta^\uparrow(1-\beta^\uparrow)$. If
a domain wall simultaneously forms in the spin-down density
profile, analogous relations hold for its current. By conservation
of the particle current, $J = j_i^\uparrow+j_i^\downarrow$, we
encounter the constraint
\begin{equation}
\alpha^\uparrow(1-\alpha^\uparrow)+\alpha^\downarrow(1-\alpha^\downarrow)
= \beta^\uparrow(1-\beta^\uparrow)+\beta^\downarrow(1-\beta^\downarrow) \, ,
\label{lb_rb_bound}
\end{equation}
which determines the IN-EX boundary.

\begin{figure}[htbp] 
\begin{center}  
\includegraphics[scale=1]{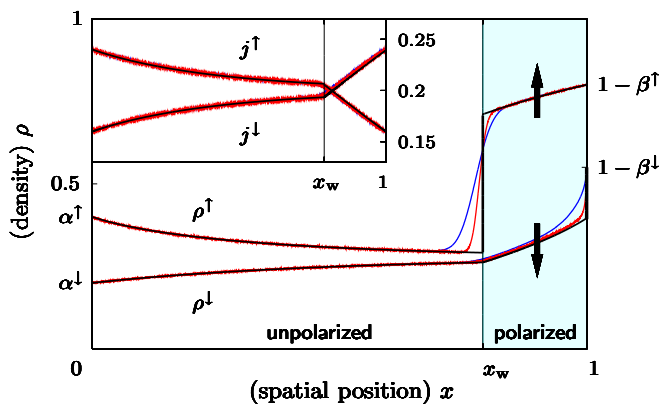}
\caption{(color online) The polarization phenomenon appearing in the IN region 
  ($\alpha^\uparrow = 0.4,\alpha^\downarrow = 0.2,\beta^\uparrow =
  0.2,\beta^\downarrow = 0.45$, and $\Omega=0.5$). A domain wall
  forms in the density profile of spin-up states, while spin-down stays in a LD phase. The spin currents, shown in the inset, are both continuous. Solid lines correspond to the
  analytical solution, while dashed lines indicate results from stochastic
  simulations on lattices with $L=2000$ (blue or dark gray) and $L=10000$
  (red or light gray).
  \label{EL}}
\end{center}
\end{figure}

Even more intriguing phase behavior emerges away from the phase
boundary between the injection and extraction dominated regions.  In
particular, we find a broad parameter regime where a localized
  domain wall forms in the density profile of one spin state, whereas
the profile of the other spin state remains almost flat. The
generic situation, as obtained from stochastic simulations and
mean-field (MF) calculations (discussed below), is exemplified in
Fig.~\ref{EL} for the  IN region.
Both spin states enter at comparable rates, and their respective
densities approach each other due to spin flips as the spins
traverse the system until they reach
 the point $x_\text{w}$.
There, the density of spin-up jumps to a high value (HD), while
the one of spin-down remains at a low level (LD). We encounter a
spontaneous polarization effect: in the vicinity of the
right boundary, the densities of spin-up and -down largely differ,
although they did not when entering the system. We will refer to
the parameter range where this spin polarization effect occurs in
the IN region as the $\text{LD-HD}_{\text{IN}}$ phase. Employing
spin-symmetry to Fig.~\ref{EL} yields a domain wall appearing in
the density profile of the spin-down state, while one concludes
from particle-hole symmetry that there is also a corresponding
$\text{LD-HD}_{\text{EX}}$ phase in the EX region. Varying the
entrance and exit rates, one can smoothly tune the domain wall
position as long as the IN-EX boundary is not crossed. If
$x_\text{w}$ passes the point $x_\text{w}=1$ in the situation of
Fig. \ref{EL}, the density of spin-up changes from the
$\text{LD-HD}_{\text{IN}}$ phase to the LD phase. On the other hand,
$x_\text{w}=0$ marks the transition between the
$\text{LD-HD}_{\text{EX}}$ and the HD phase. Regarding the domain wall
position $x_\text{w}$ as an order parameter, these transitions are
continuous.

The formation of a localized domain wall can be understood from the
continuity of the spin currents (Fig. \ref{EL}), which in turn arises from the only
weak leakage current in the mesoscopic limit \cite{parmeggiani-2003-90}.
Say the domain wall forms in the spin-up state.
Then
we have to match the currents
adjacent to the left ($l$) and right
($r$) of this wall,
which in the limit of large system size are given
by $j^\uparrow_{l/r}(x) = \rho^\uparrow_{l/r}(x)
[1-\rho^\uparrow_{l/r}(x)]$.
Both currents should coincide at the
position of the domain wall, $j^\uparrow_l (x_\text{w}) = j^\uparrow_r
(x_\text{w})$, while the density shows a discontinuity. Together, we
arrive at the condition $\rho^\uparrow_r (x_\text{w}) = 1 -
\rho^\uparrow_l (x_\text{w})$ for the domain wall position $x_\text{w}$.  The spatial
dependence of the density implies that a suitable position
$x_\text{w}$ indeed exists within a certain parameter region.

Let us now underpin the so far general discussion by a quantitative
analytical description. Consider therefore the average density
$\rho_i^{\uparrow, \downarrow}$ at site $i$ for spin-up and -down
states. The dynamical rules yield equations for their time evolution,
which upon factorizing two-point correlations reduce to a closed set
of difference equations;  such MF approximations have been fruitfully applied 
within related contexts \cite{parmeggiani-2003-90, derrida-1992-69,evans-1995-74}.
In a continuum limit, the difference equations turn into differential   
equations, which to leading order in the lattice constant $1/L$ take
the form     
\begin{subequations}
\begin{equation}
 (2\rho^\uparrow-1) \partial_x\rho^\uparrow  +
 \Omega\rho^\downarrow - \Omega\rho^\uparrow = 0 \, ,
\end{equation}
\vspace{-0.9cm}
\begin{equation}
 (2\rho^\downarrow-1) \partial_x\rho^\downarrow  +
 \Omega\rho^\uparrow - \Omega\rho^\downarrow = 0 \, .
 \end{equation}
 \label{cont}
\end{subequations}
Together   
with appropriate boundary conditions arising from the entrance and
exit processes, they allow a straightforward analytic solution,
the details of which will be presented in a forthcoming
publication \cite{reichenbach-long}.  We have compared the
analytic solution to extensive stochastic simulations through
careful finite-size scaling analysis. Upon increasing the system
size $L$, the densities converge to the analytical prediction, as
exemplified in Fig.~\ref{EL}. The observed exactness of the
analytical density profiles in the limit of large systems originates,
on the one hand, in the exact mean-field current-density relation in the TASEP \cite{derrida-1998-301}. 
On the other hand, the coupling of the two internal states in our model locally tends to zero when the system
size $L$ is increased, such that correlations between them are washed
out. The situation is somewhat analogous to
TASEP combined with Langmuir kinetics \cite{parmeggiani-2003-90}.
  
The analytical approach therefore allows one to obtain the exact phase
diagrams of the system. In particular, we may determine the regions
where phase separation and thus the polarization phenomenon occurs.
In general, a large variety of different phases appears with
discontinuous as well as continuous transitions between them. As
anticipated by our symmetry arguments, a discontinuous transition
accompanied by a domain wall delocalization occurs across the border
between the injection and extraction dominated regions, which is given
by Eq.~\eqref{lb_rb_bound}.  Continuous transitions appear inside the
IN as well as the EX region, when the
domain wall leaves the system at one of its boundaries: $x_w=0$ and $x_w=1$. 
The locations in phase space where these special domain wall positions occur,
the phase boundaries, can be obtained from
the analytical solution of the density profiles, as the latter reveals
the position $x_w$; details will be presented elsewhere \cite{reichenbach-long}.

\begin{figure}
\begin{center}
\includegraphics[scale=1]{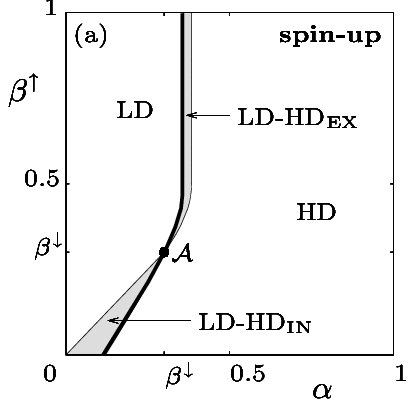}
\hspace{0.15cm}
\includegraphics[scale=1]{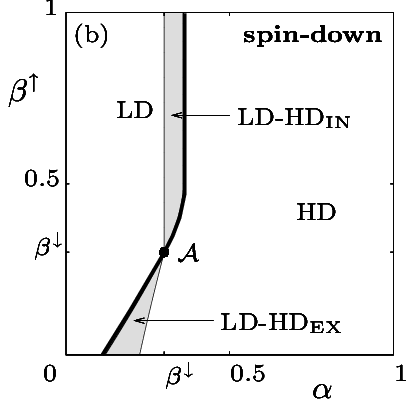}
\caption{Phase diagram for equal injection rates
  $\alpha^\uparrow=\alpha^\downarrow\equiv\alpha$. Phases for spin-up
  and -down are shown in panels (a) and (b),
  respectively, as a function of $\alpha$ and $\beta^\uparrow$ for
  fixed values $\beta^\downarrow=0.3$ and $\Omega=0.15$. Phase
  separation (shaded areas) arises in the IN as well as the
  EX region.  The delocalization-transition line (thick line) is  identical in (a) and (b).
 Thin lines correspond to continuous transitions.
\label{phasediag_equal_alpha}}
\end{center}
\end{figure}

For illustration, we  focus on the special case of equal entrance
rates, $\alpha^\uparrow = \alpha^\downarrow \equiv \alpha$, which
already exhibits the main features of the possible non-equilibrium
steady states. In particular, when a domain wall occurs in the
density profile of one of the spin states within the
IN region, it lucidly shows the polarization phenomenon:
both spins have equal densities in the vicinity of the left
boundary, but strongly differ at the right one.  The phase diagram
for the spin-up and spin-down states, resulting from the analytical
solution,  is presented in Fig.~\ref{phasediag_equal_alpha}.
In the two-lane interpretation of the model, the states refer to
the upper and lower lane. For 
both internal states the same first order line marks the
transition between the IN and EX region. The
polarized state (shaded area), where a domain wall appears either
in the spin-up or -down state, intervenes between the LD and HD
phases. There are continuous transitions from  pure (LD, HD) to
coexistence phases as the domain wall enters or exits the system at
the boundary.
 The lines marking these transitions intersect the IN-EX boundary in a
multicritical point $\mathcal{A}$, where all boundary rates
 equalize, $\alpha = \beta^\uparrow = \beta^\downarrow$.

 \vspace{-0.2cm}
Consider now a horizontal path through the phase diagram below the
multicritical point $\mathcal{A}$.  For small values of the
injection rate $\alpha$ the system is in the injection dominated
region, and both spin states (lanes) are in a homogeneous LD
state. Then, upon crossing the phase boundary for the spin-up
state, the system switches to a polarized state, similar to
Fig.~\ref{EL}, such that a domain wall appears in the density
profile of the spin-up state (upper lane), entering continuously
from the right boundary; for the same parameter range the spin
down state stays in a homogeneous LD state.  Approaching the
IN-EX boundary this domain wall delocalizes, and upon
crossing relocalizes again, but now as a domain wall in the spin-down
state (lower lane). As the spin-up state has turned to a HD
phase, we encounter  polarization near the left boundary.
Crossing the IN-EX boundary, the system
thus switches its polarization from the right boundary to the left one.
 A further
increase of the injection rate finally shifts the position of the
domain wall to the left  boundary such that the system  remains
in a HD phase for both spin states (lanes).  For a path through
the phase diagram above $\mathcal{A}$, similar arguments hold.

Two of the lines that mark continuous transitions are readily
obtained.  The transition from the LD to the $\text{LD-HD}_{\text{IN}}$
phase  is determined by $x_\text{w} = 1$; since $\rho^\uparrow(x)
= \rho^\downarrow(x) = \alpha$ for $x < x_\text{w}$, it is located
along the diagonal, $\alpha = \beta^\uparrow$, for  spin-up state and parallel to the vertical axis, $\alpha = \beta^\downarrow$,
for the spin-down state. We emphasize that these phase boundaries in the
injection dominated regime do not depend on the magnitude of the
gross spin flip rate $\Omega$; i.e., qualitatively tuning the
system's state is possible only upon changing the injection or
extraction rates. The lines corresponding to continuous transition
in the EX region are more
involved~\cite{reichenbach-long}. Its most notable feature is that
the width of the polarized phase decreases with increasing spin
flip rate $\Omega$, until it finally vanishes in the limit $\Omega \to
\infty$.

\acknowledgments The authors are grateful for
insightful discussions with Felix von Oppen and Ulrich Schollw\"ock.


\end{document}